\documentclass[fleqn,12pt,twoside]{article}
\usepackage{espcrc1}

\usepackage{graphicx}
\usepackage[figuresright]{rotating}

\newcommand{\msun}{$M_{\odot}$}
\newcommand{\osi}{$^{16}${O}}
\newcommand{\ose}{$^{17}${O}}
\newcommand{\oei}{$^{18}${O}}

\newcommand{\AmS}{{\protect\the\textfont2
  A\kern-.1667em\lower.5ex\hbox{M}\kern-.125emS}}

\title{HR4049: signature of nova nucleosynthesis ?}

\author{M. Lugaro\address[IoA]{Institute of Astronomy, University of Cambridge\\
Madingley Road, Cambridge CB3 0HA, UK},
O. Pols\address{Astronomical Institute, University Of Utrecht\\
P.O. Box 80000, 3508 TA Utrecht, The Netherlands},
A.I. Karakas\address{Institute for Computational Astrophysics,
Department of Astronomy \& Physics \\ Saint Mary's University, Halifax, Canada} and 
C.A. Tout\addressmark[IoA]}

\begin{document}

\maketitle

\begin{abstract}

The post-Asymptotic Giant Branch (AGB) star HR4049 is in an eccentric binary
system with a relatively short period probably surrounded by a
dusty circumbinary disk. Extremely anomalous oxygen isotopic ratios, 
\osi/\ose~ $\simeq$ \osi/\oei~ $\simeq$ 7, have been 
measured from CO$_2$ molecules likely residing in the disk. 
Such a composition cannot be explained in the framework of AGB and post-AGB
evolution while it can be qualitatively associated with the nucleosynthesis
occurring during nova outbursts. We discuss nova models, the presence of a
white dwarf companion to HR4049 and possible scenarios for the dynamical
evolution of this binary system.
Circumbinary disks in which mixing occurs between red-giant and nova material may also be 
invoked as the site of formation of some rare types of meteoritic presolar grains.

\end{abstract}

\section{Introduction}

\begin{figure}[htb]
\begin{minipage}[t]{80mm}
\includegraphics[width=79mm]{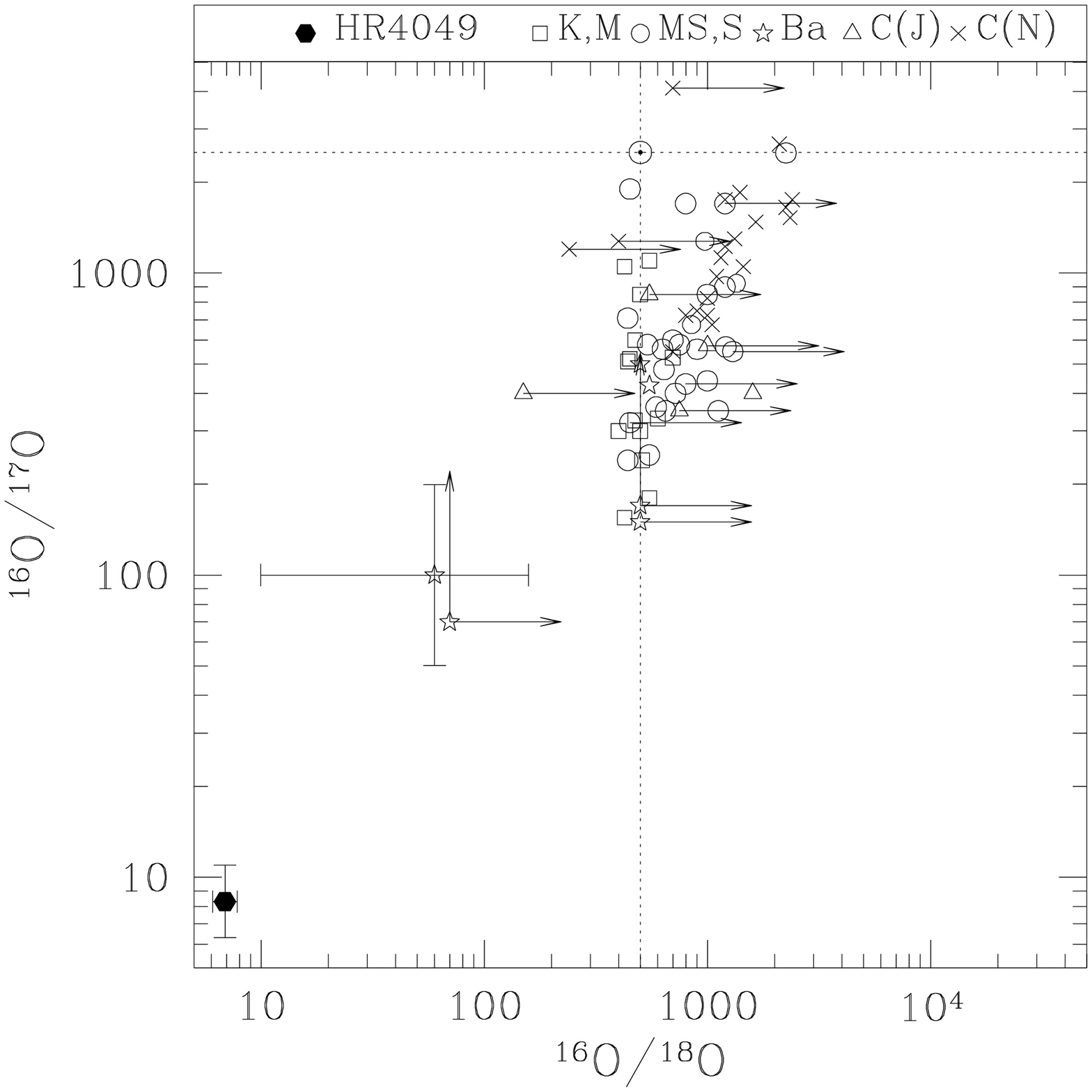}
\caption{Oxygen isotopic ratios in HR4049 are compared to those of other evolved 
stars.}
\label{fig:oxygen}
\end{minipage}
\hspace{\fill}
\begin{minipage}[t]{75mm}
\includegraphics[width=79mm]{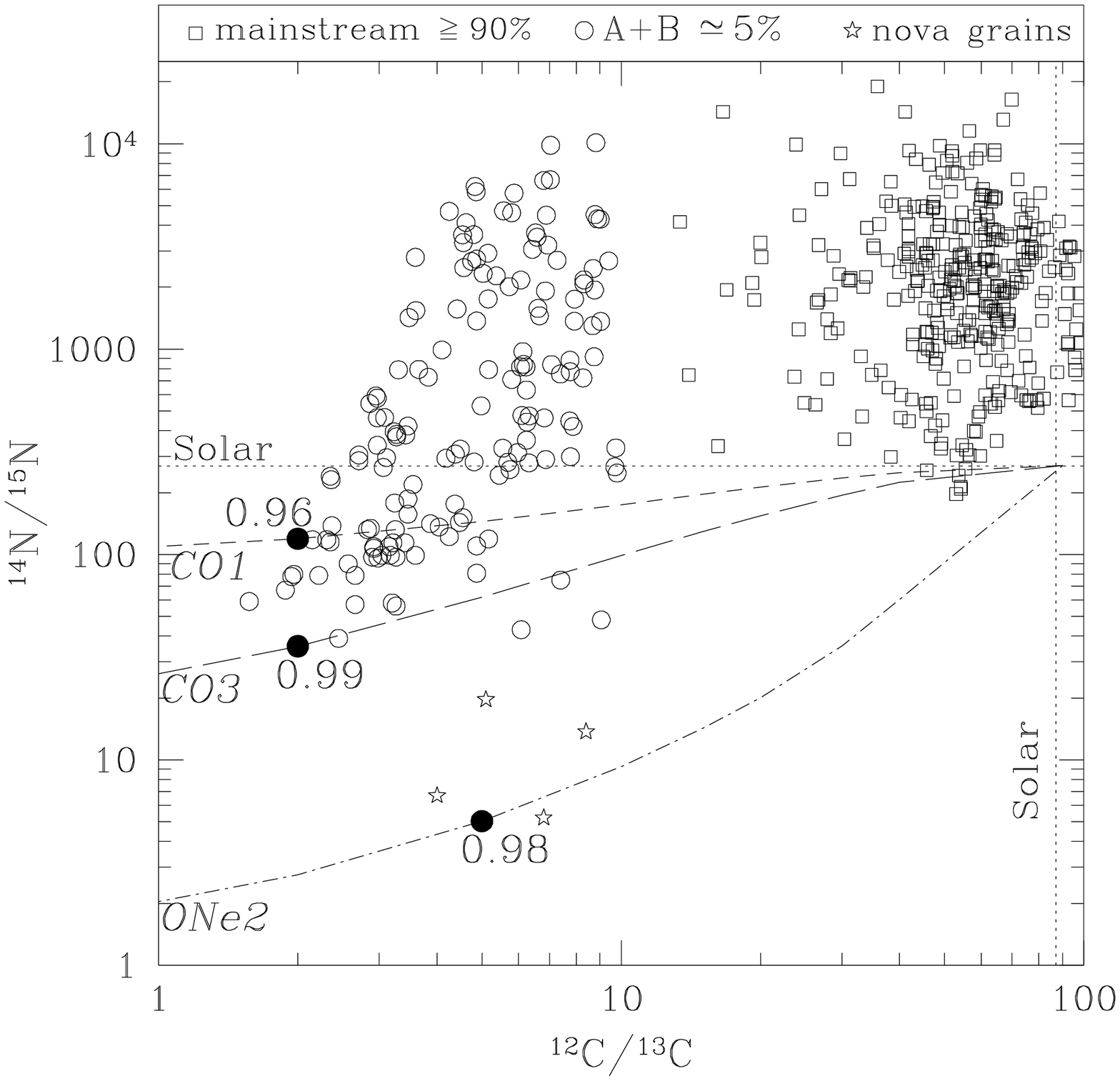}
\caption{C and N isotopic ratios in mainstream, A+B and nova SiC grains. Lines 
indicates mixing of solar with nova material from models by \cite{jose:98}. Numbers 
along the lines indicates the mass fraction of solar material in the mix.}
\label{fig:grains}
\end{minipage}
\end{figure}

HR4049 is a post-AGB star member of a binary system with period of 429 days and high 
eccentricity, $\epsilon$=0.3. HR4049 is the prototype of a group of post-AGB stars in binary 
systems showing a high degree of metal depletion, $< -$ 4 dex \cite{vanwinckel:95}. The key 
to this feature is probably the presence of a massive, stable, long-lived and dusty
circumbinary disk: after dust forms in the disk the gas depleted of refractory 
elements is re-accreted by the star, see e.g. \cite{dominik:03}. Oxygen isotopic 
ratios have been measured from 
CO$_2$ molecules likely residing in the disk \cite{cami:01}. These ratios are extremely 
anomalous: $^{16}$O/$^{17}$O = 8.3 $\pm$ 2.3 and $^{16}$O/$^{18}$O = 6.9 $\pm$ 0.9, as 
illustrated in Figure \ref{fig:oxygen} they are one to more than two orders of magnitude 
lower than observed in any evolved star! Such a composition cannot be produced in AGB 
and post-AGB stars because their composition is characterised by the effect 
of He burning, during which \ose~ and \oei~ are destroyed, and proton-capture processes 
during which \ose~ is produced, but not to the level shown by HR4049, and \oei~ is 
destroyed.

\section{Nova nucleosynthesis ?}

If the companion of HR4049 is a white dwarf (WD), accretion from HR4049 could have
triggered nova outbursts in the same way as predicted \cite{iben:96} and observed
\cite{belczynski:98} in some symbiotic systems. The nova ejecta could have been partly
intercepted by HR4049 or the circumbinary disk, to produce the observed anomalous
oxygen ratios. The mass of HR4049 is between 0.56 -- 0.67 \msun~ and the mass function
of the system is $f(m)$=0.158 $\pm$ 0.004. The estimated inclination is $\sim$ 60
degrees \cite{dominik:03}. Hence the mass of the companion can be estimated to be 
around
0.75 -- 0.82 \msun, which indicates we should look into nova models with a CO white
dwarf. Oxygen isotopic ratios for CO nova models are shown in Table~\ref{table:1}.
Results depend on several parameters: the mass, temperature and luminosity of the WD,
the accretion rate, the composition of the envelope produced by the accreted material
and the level of mixing of WD material into this envelope. In order to be able to match
the observed ratio \osi/\ose~ $\leq$~8 is needed. Nova models appear to allow such
composition, depending on the choice of parameters.  Note also that variations of the
\osi($p,\gamma$), \ose($p,\gamma$) and \ose($p,\alpha$) reaction rates within their
uncertainties give rise to \ose~ abundances changes by factor $\leq$ 30
\cite{iliadis:02}.

\begin{table}[htb] 
\caption{Ranges of oxygen isotopic ratios from nova models}
\label{table:1} 
\newcommand{\m}{\hphantom{$-$}}
\newcommand{\cc}[1]{\multicolumn{1}{c}{#1}} \renewcommand{\tabcolsep}{2pc} 
\renewcommand{\arraystretch}{1.2} 
\begin{tabular}{@{}llll} 
\hline

Reference & $M_{WD}$ (\msun) & \osi/\ose & Varied parameters \\

\hline

\cite{starrfield:72} & 1. & 2.6 --- 21 & Luminosity and $^{12}$C \\

\cite{starrfield:78} & 1. & 0.52 --- 120 & Envelope mass \\
& & & and composition \\

\cite{kovetz:97} & 0.65 & 19 --- 52 & WD temperature and \\
& 1. & 3.1 --- 17 & the accretion rate \\

\cite{jose:98} & 0.8 & 39 --- 57 & Level of mixing \\
& 1. & 30 & with core material \\

 &  & $^{16}O/^{18}O$ &  \\

\cite{jose:98} & 0.8 & {\it 155 --- 446} & Level of mixing \\
& 1. & {\it 109} & with core material \\

\hline
\end{tabular}\\[2pt]
\end{table}

Isotopic ratios of \osi/\oei~ presented by Ref.~\cite{jose:98} for CO nova models 
are also shown in Table~\ref{table:1}. These nova models do not 
produce the observed ratio, requiring \osi/\oei~ $\leq$ 7. However, the final 
abundance of \oei~ originates predominantly from the decay
of $^{18}$F, whose abundance is very sensitive to $^{18}$F($p,\alpha$), 
\ose($p,\gamma$) and \ose($p,\alpha$) reaction-rate variations. Within rates 
uncertainties, the $^{18}$F abundance changes by factors of $\leq$ 100 
\cite{iliadis:02}. Thus, within current uncertainties also the required \osi/\oei~ 
ratio could be produced by CO nova models.


The main question in relation to the nova scenario is whether the companion of HR4049
is really a WD.
There is no sign of the presence of a WD in the ultraviolet spectrum so that, if 
there is one, its temperature must be lower than $\sim$ 20,000 K \cite{monier:99}.
If the WD is still accreting from HR4049 winds it could emit X-rays. 
ROSAT observations suggest that there might be a marginal 
detection about 20'' away from HR4049 position, on the edge of being 
compatible with the positional accuracy of ROSAT (F. Verbunt, personal
communication). A more sensitive observational test can be obtained making use of the
higher resolution of CHANDRA.

Many other issues are open with regards to the evolution of the system. With
a separation of 190 R$_{\odot}$, the Roche-lobe radius is $\sim$ 
76 R$_{\odot}$. The radius of HR4049 at the tip of the AGB would have been 
$\sim$ 250 R$_{\odot}$, much larger than the Roche-lobe radius. Several
scenarios can be proposed for the evolution of this binary system.
For example, if the system initially had a larger separation nova
outbursts could have been triggered by wind accretion before
dynamically unstable Roche-lobe overflow resulted in common envelope
evolution and orbital shrinkage. Alternately, if the mass ratio
reached a value smaller than about 2/3 the system could have undergone
stable Roche-lobe overflow for some time and triggered nova
outbursts. However, in either case it is difficult to account for the
current high eccentricity of the binary system.

\section{Relevance to meteoritic presolar grains}

The compositions of a few presolar SiC and graphite grains indicate a 
nova origin \cite{amari:01a}. There a few open problems and 
questions related to this origin: (i) mixing between nova and more normal material is 
needed, (ii) C/O$>$1 is not typically produced by novae, but is a necessary condition to 
condensate SiC and graphite, and (iii) grain compositions indicate an ONe nova  
origin, but where are the grains from CO novae? 

If we consider as a site of grain formation the circumbinary disks around evolved 
stars, such as that around HR4049, we can find solutions to the three problems listed above.
(i) In a circumbinary disk nova material could be mixed with red giant 
material. (ii) Material from the red giant stars could be carbon rich and, finally, as 
illustrated in Figure \ref{fig:grains},
(iii) grains originating from material affected by CO nova nucleosynthesis could be 
found among grains belonging to the A and B SiC populations, whose composition has 
so far remained unexplained \cite{amari:01b}. 
An objection to this scenario is that there are few
circumbinary disks around evolved stars and not all of them could
have experienced nova nucleosynthesis. How can they produce a
significant fraction (few percent) of the recovered meteoritic
presolar SiC grains ? A possible explanation is that 
many presolar meteoritic SiC grains, which have relatively large sizes up to a few 
$\mu$m, could have been {\it preferentially} produced in circumbinary disks,
rather than in single-star outflows \cite{jura:97}.

\section{Summary and conclusions}
The O composition of HR4049 could 
represent the first time that isotopic ratios showing the signature of nova 
nucleosynthesis are measured in a stellar environment.
This composition could have
resulted from nova outbursts if: (i) the companion of HR4049 is a WD and
(ii) the $^{18}$F($p,\alpha$) rate is close to its lower limit, so that
\oei~ can be produced to the same level as \ose. Dedicated nova models are needed 
to address the issue. Many problems related to the system 
dynamics are still open. A significant fraction of presolar SiC grains could have 
originated in circumbinary disks with material being affected by nova
nucleosynthesis.

\end{document}